\documentclass[conference]{IEEEtran}
\IEEEoverridecommandlockouts
\usepackage{cite}
\usepackage{amsmath,amssymb,amsfonts}
\usepackage{graphicx}
\usepackage{textcomp}
\usepackage{xcolor}

\usepackage{booktabs} %
\usepackage{csvsimple}
\usepackage{algorithm}
\usepackage{algpseudocode}
\usepackage{mathtools}
\usepackage{hyperref}

\def\BibTeX{{\rm B\kern-.05em{\sc i\kern-.025em b}\kern-.08em
    T\kern-.1667em\lower.7ex\hbox{E}\kern-.125emX}}

\usepackage{tikz}

\newcommand\copyrighttext{%
  \footnotesize \textcopyright 2022 IEEE. Personal use of this material is permitted.
  Permission from IEEE must be obtained for all other uses, in any current or future
  media, including reprinting/republishing this material for advertising or promotional
  purposes, creating new collective works, for resale or redistribution to servers or
  lists, or reuse of any copyrighted component of this work in other works.
  }
\newcommand\copyrightnotice{%
\begin{tikzpicture}[remember picture,overlay]
\node[anchor=south,yshift=10pt] at (current page.south) {\fbox{\parbox{\dimexpr\textwidth-\fboxsep-\fboxrule\relax}{\copyrighttext}}};
\end{tikzpicture}%
}

\begin{document}

\title{Efficient Runtime Profiling for Black-box Machine Learning Services on Sensor Streams}

\author{
\IEEEauthorblockN{Soeren Becker, Dominik Scheinert, Florian Schmidt, and Odej Kao}
\IEEEauthorblockA{Technische Universit{\"a}t Berlin, Germany, \{firstname.lastname\}@tu-berlin.de}
}

\maketitle
\IEEEpubidadjcol
\copyrightnotice

\begin{abstract}
In highly distributed environments such as cloud, edge and fog computing, the application of machine learning for automating and optimizing processes is on the rise.
Machine learning jobs are frequently applied in streaming conditions, where models are used to analyze data streams originating from e.g. video streams or sensory data.
Often the results for particular data samples need to be provided in time before the arrival of next data.
Thus, enough resources must be provided to ensure the just-in-time processing for the specific data stream.

This paper focuses on proposing a runtime modeling strategy for containerized machine learning jobs, which enables the optimization and adaptive adjustment of resources per job and component.
Our black-box approach assembles multiple techniques into an efficient runtime profiling method, while making no assumptions about underlying hardware, data streams, or applied machine learning jobs.
The results show that our method is able to capture the general runtime behaviour of different machine learning jobs already after a short profiling phase. 
\end{abstract}

\begin{IEEEkeywords}
Profiling, Runtime Prediction, Resource Estimation, Performance Modeling, Heterogeneous Environments
\end{IEEEkeywords}

\section{Introduction}
\label{sec:introduction}

The ever increasing scale and distribution of cloud environments not only enables new use cases in the area of Industry 4.0, smart health care, mobile computing, or Internet of Things in general, but also increases the amount of generated data. 
Especially in the aforementioned domains, different types of sensors are employed to collect data in order to represent real-world phenomena, enable technologies such as digital twins and improve the environmental awareness. 
Moreover, the combination and analysis of sensor streams allows for new usage scenarios and business models, providing e.g. predictive maintenance services, healthcare monitoring or remote machine orchestration.

Besides the advances in data analysis through i.e. sophisticated deep-learning models and the increase in data acquisition, the locality of data processing tasks is also shifted closer to the actual sources: Upcoming technologies such as edge and fog computing enable the execution of machine learning (ML) models on small and lightweight devices, which can be located across e.g. a smart city. 
Consequently, the amount of heterogeneous devices in terms of hardware components (i.e. different CPU architectures, CPU cores, memory or available accelerators such as GPUs) in
the edge cloud continuum is growing rapidly. 

Although lightweight virtualization technologies such as containerization, in combination with orchestration frameworks like Kubernetes, simplified the deployment of analysis tasks in these
environments, considerable challenges still remain. 
One of the main benefits of analyzing the (sensor) data streams as close as possible to the data source, is the accelerated response time for situations indicated in the data. 
In order to act as soon as possible on e.g. outliers detected in the data stream, enough resources for the ML jobs need to be provided to ensure a just-in-time computation of the ingested data. 
Consequently, this results in strict requirements for hardware resources to permit the analysis of a sample in the data stream before the next sample arrives. 
Especially in settings of high-frequent data streams, the bounds get increasingly important.

Hence, a runtime model of the applied stream-based ML jobs under different resource limitations is needed to determine the required resources for just-in-time computation with given sample frequencies in the data stream. 
Runtime prediction in the context of cloud environments and big data is often based on historical data of job executions and involves domain knowledge about the data and applied algorithms, static runtime targets and employed hardware~\cite{thamsenSelectingResourcesDistributed2016,scheinertEnelContextAwareDynamic2021}.

Considering the heterogeneous nature of the previously mentioned highly distributed environments, a global runtime model of a job might not be fitting for each available device type in the infrastructure. 
In addition, new sensors can be connected in an ad-hoc manner to available devices, yielding further data streams and resulting in stream-based analytic jobs for which historical data is not yet available. 
Finally, the sample frequency in the data stream can vary over time or configuration which motivates an efficient approach to model the runtime behaviour of jobs directly on the respective and possibly lightweight devices.

Therefore, we aim to investigate the problem of runtime profiling and prediction for black-box streaming ML jobs. 
In this paper, we propose a profiling approach which is executed on the respective devices each time a new job is started. 
Furthermore, we present several selection strategies for profiling points and introduce the concept of synthetic runtime targets. 
The generated model can be used for the adaptive adjustment of resources to set the highest restriction of resources, while still meeting runtime targets of the incoming data.

Summarizing, as key contributions of this paper we
\begin{itemize}
    \item propose an efficient runtime profiling approach in order to enable the adaptive adjustment of resources
    \item present and implement a nested modeling strategy as well as appropriate alternatives to find suitable profiling points while simultaneously optimizing the overall profiling time 
    \item conduct an extensive evaluation of the proposed profiling and selection strategies across seven different machine types and several machine learning algorithms 
\end{itemize}

\textit{Outline:} The reminder of the paper is structured as follows: \autoref{sec:related_work} discusses the related work for runtime profiling and fingerprinting in the context of ML jobs. 
\autoref{sec:approach} carries on with the description of our efficient runtime profiling method, while~\autoref{sec:evaluation} presents the evaluation of the proposed method across a set of heterogeneous nodes and compares its results with comparative methods. 
Finally,~\autoref{sec:conclusion} concludes the paper and outlines possible future work.
\section{Approach}
\label{sec:approach}
\begin{figure*}[ht]
    \centering
    \includegraphics{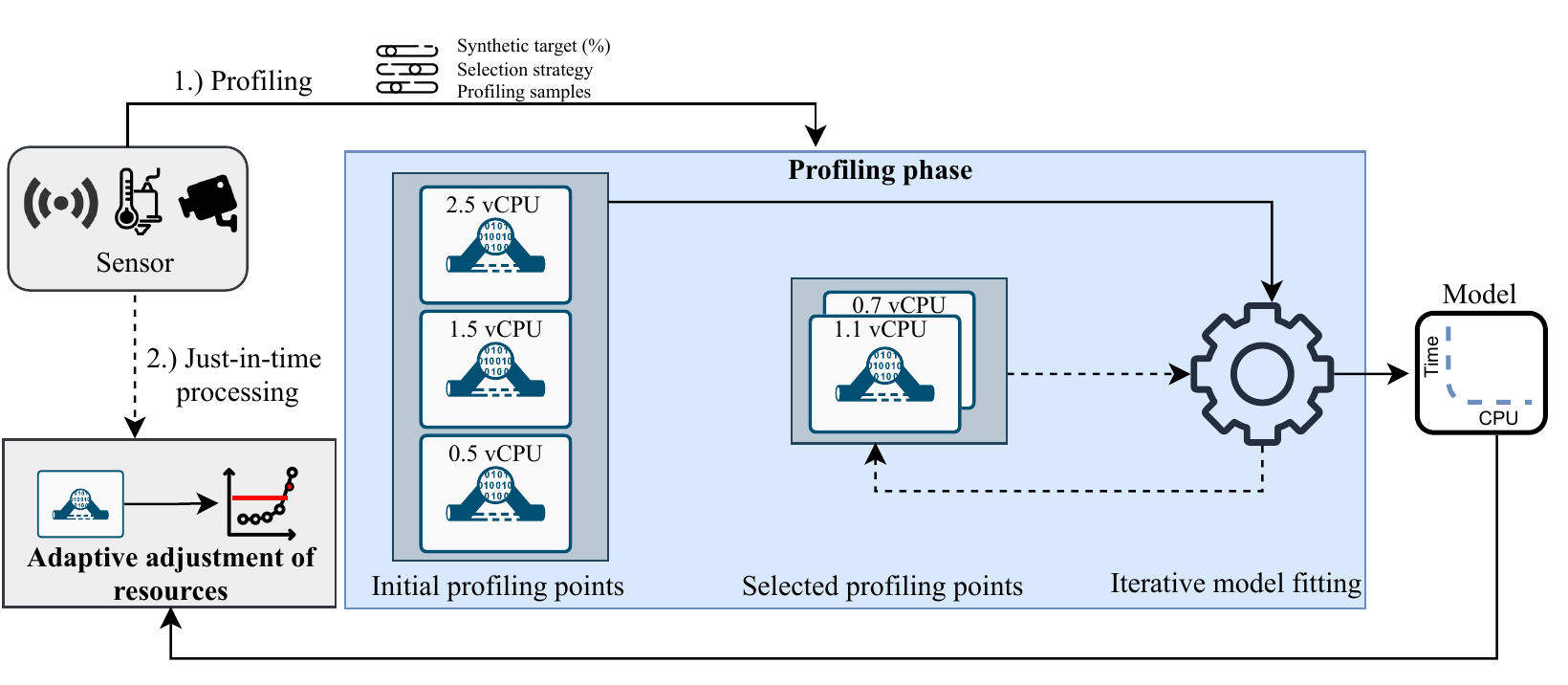}
    \caption{An abstract description of our approach. Newly generated data streams are first analyzed during a profiling phase in order to generate a runtime model which is subsequently used for the adaptive adjustment of resources in order to enable a just-in-time processing of incoming data samples.}
    \label{fig:approach}
\end{figure*}
This section presents our solution to obtaining a general runtime model for black-box machine learning services.
To this end, as depicted in \autoref{fig:approach}, we propose an approach that combines both parallel and iterative profiling runs together with an iterative replacement strategy for the employed runtime model, such that for any number of investigated resource configurations, a more suitable function is fitted. Finally, the resulting model can be used to dynamically adjust the resources of analysis jobs using i.e. Kubernetes, in order to enable a just-in-time processing of incoming data samples.

\subsection{Runtime Model}
Since profiling of applications induces additional costs, we seek a lightweight solution for both profiling and modeling.
In~\cite{gulenko2020bitflow}, the authors propose an approximation function for modeling the computation time with respect to limited resources $R$.
As shown in~\cite{Becker2020aiopsedge}, it is reasonable for certain stream-based ML jobs to focus on CPU only, as memory overhead is oftentimes constant.
We design an adapted version of this function $compute(\cdot)$ and define it as:
\begin{equation}
\label{eq:base}
compute(R) = a \cdot (R \cdot d)^{-b}+c
\end{equation}
This equation includes four parameters to be optimized, namely $a,b,c$, and $d$. 
Previous works~\cite{gulenko2020bitflow,Becker2020aiopsedge,beckerLocalOptimisticSchedulingPeriodic2021a} found that learning these four parameters from a profiling phase can provide high quality profiling approximations for stream-based machine learning jobs. 
Due to the four parameters, fitting a curve is just possible with at least five points from a profiling phase. 
However, it remains unclear how to best proceed when no data points are available yet, and how to select valuable next data points based on a handful of already known.
We deliver an answer to the question of limited data points by proposing an iterative replacement strategy for the employed function to fit. 
Formally, let $|R|$ denote the number of data points, we employ a function to fit as:

\[f(R)= \begin{cases*}
    R^{-1}, & if  $|R|=1$ \\
    a \cdot R^{-1}, & if $|R|=2$ \\
    a \cdot R^{-b}, & if $|R|=3$ \\
    a \cdot R^{-b} + c, & if $|R|=4$ \\
    a \cdot (R \cdot d)^{-b} + c, & else
    \end{cases*}\]

In other words, we choose our fitting function dependent on the number of available data points.
We thus incrementally improve our prediction capabilities with each newly profiled data point until~\autoref{eq:base} can be employed. 

\subsection{Selection Strategy}
\label{sec:selection_strategy}
Given an envisioned target runtime, our approach to runtime modeling can be used to predict the next CPU limitation to investigate which most likely better matches the runtime target.
Over time, with an increasing number of data points and also more sophisticated fitting functions usable, the runtime behavior of a black-box machine learning service with respect to its CPU limitations can be sufficiently approximated. 
Yet, this procedure is vulnerable in two aspects:
\begin{itemize}
    \item The learned runtime behavior is highly dependent on the concrete runtime target, which means that a fitted function does not necessarily generalize well, i.e. its interpolation and extrapolation capabilities are potentially limited.
    \item As the runtime increases exponentially for small limitations imposed on the CPU, a disadvantageous selection strategy can significantly prolong the profiling phase.
\end{itemize}
To ensure that the exponential relationship between CPU limitations and runtimes is appropriately captured, we introduce the idea of \emph{synthetic} targets, i.e. we target the runtime corresponding to a CPU limitation that is just small enough to guarantee an inspection of most of the exponential curve by our runtime modeling approach.
By profiling this small CPU limitation and using the observed runtime as runtime target, all subsequent CPU limitations will be selected accordingly.
Moreover, since the profiling of especially small CPU limitations takes a comparably long time while only utilizing a fraction of the overall available resources, we initially conduct multiple profiling runs in \emph{parallel} and ensure that the range of CPU limitations to investigate is appropriately covered.
This effectively reduces the overall profiling time and makes use of the available resources efficiently. 
Formally, given a set of possible CPU limitations $L=\{l_{\min}, l_{\min} + \delta, \ldots, l_{\max} - \delta, l_{\max}\}$ with lower bound $l_{\min}$, upper bound $l_{\max}$, and logical step size $\delta$, we aim at running $n$ initial profiling runs in parallel such that the selected CPU limitations $R_{\text{initial}}$ are unique, sum up to at most $l_{\max}$, and sufficiently represent the range of possible CPU limitations while minimizing the profiling time:
\begin{equation}
    R_{\text{initial}} \in \{(c_1, \ldots, c_n)| c_i\in L, c_1 + \ldots c_n \le l_{\max}\}
\end{equation}
In our evaluation, we investigate multiple sets $R_{\text{initial}}$ and work out their impact upon profiling and modeling.
    
\subsection{Early Stopping}
\label{sec:earlystopping}
In order to provide an efficient profiling, it is crucial to run the profiling for a certain period of time during single limitations, but the duration should be as short as possible in order to limit the overall time. 
The time is reported by the model itself for each single computed sample. 
The main task is therefore to determine when enough samples have been reviewed to be confident in the computation time per sample, in order to stop the profiling of the given resource limitation.

\begin{figure}
    \centering
    \includegraphics{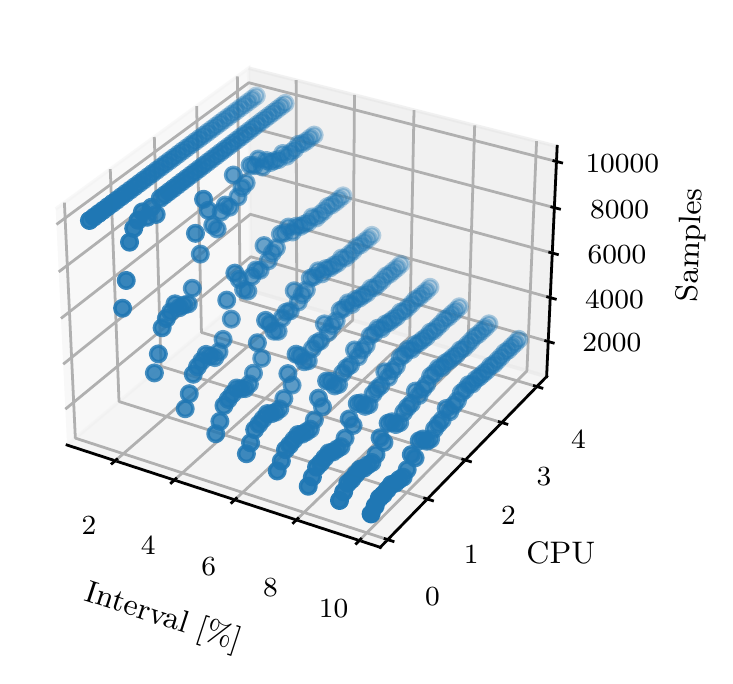}
    \caption{Early stopping results for the LSTM algorithm on a Raspberry Pi 4 with a 95\% confidence interval.}
    \label{fig:earlystopping}
\end{figure}

We propose to apply t-distribution to determine the confidence interval with a user defined percentage of confidence (typically 95\% or 99.5\%). T-distribution assumes Gaussian distribution of the underlying signal, but also includes the confidence of number of samples seen so far. 
With more data, the confidence increases, but is highly dependent on the variation of values. 
With a given confidence interval, the execution is going to stop after finite time since the size of the interval is used as stopping criteria. 
This is depicted in~\autoref{fig:earlystopping} and holds true for any concrete CPU utilization.
For example, assuming that we want our confidence interval $CI=[a,b]$ with boundaries $a$ and $b$ computed at the 95\% confidence level to be smaller than a specific fraction $\lambda\in (0, 1)$ of the empirical mean $\bar{X}$, i.e. $\vert b-a\vert < \lambda \cdot \bar{X}$, it is required to profile more samples with a fraction of 2\% as it would be the case for 10\%.

\section{Evaluation}
\label{sec:evaluation}
This section presents our experiment setup, the utilized datasets, our exhaustive comparison of methods, and discussion of the results.

\subsection{Experiment Setup}
To evaluate our approach across several devices, we created and collected datasets from the devices listed in~\autoref{tab:devices}.

As three example workloads, we implemented \emph{Arima}, \emph{Birch} and \emph{LSTM}-based anomaly detection algorithms in the IFTM framework \cite{schmidt2018iftm} which allows for online and unsupervised outlier detection in data streams. Therefore, they are a good use case for a generalized runtime model to enable dynamic adjustments of resource limitations in case of i.e. changing sample arrival rates in the analyzed data streams.

\begin{table}[ht]
\centering
\caption{Hardware Specifications}
\label{tab:devices}
\begin{tabular}{@{}lcrr@{}}
\toprule
Hostname & Type                                                            & CPU cores & Memory  \\ \midrule
wally & \begin{tabular}[c]{@{}c@{}}Commodity server \\ (Intel Xeon E3-1230)\end{tabular} & 8 & 16 GB \\ \midrule
asok  & \begin{tabular}[c]{@{}c@{}}Commodity server\\  (Intel Xeon X5355)\end{tabular}   & 8 & 32 GB \\ \midrule
pi4      & Raspberry Pi 4B                                                 & 4         & 2  GB   \\ \midrule
e2high   & \begin{tabular}[c]{@{}c@{}}GCP VM \\ (e2-highcpu)\end{tabular}  & 2 vCPU    & 2 GB    \\ \midrule
e2small  & \begin{tabular}[c]{@{}c@{}}GCP VM \\ (e2-small)\end{tabular}    & 2 vCPU    & 2 GB    \\ \midrule
e216     & \begin{tabular}[c]{@{}c@{}}GCP VM \\ (e2-highcpu)\end{tabular}  & 16 vCPU   & 16 GB   \\ \midrule
n1       & \begin{tabular}[c]{@{}c@{}}GCP VM \\ (n1-standard)\end{tabular} & 1 vCPU    & 3,75 GB \\ \bottomrule
\end{tabular}
\end{table}

\paragraph{Data Acquisition}
We provided the aforementioned algorithms in docker containers on the respective nodes and utilized a dataset of 10,000 samples with 28 monitoring metrics as example data stream.
During the data acquisition phase we leveraged the Docker execution engine to limit the CPU utilization of running containers: For each algorithm we started with all available vCPUs on the nodes and used the dataset as an input for the respective algorithm. 
As metrics we measured the average processing time per sample and subsequently decreased the allocated vCPUs by 0.1 for each following execution.

In the following experiments, the accumulated results were used in order to evaluate our approach in conjunction with multiple concrete selection strategies.

\paragraph{Selection Strategies} 
In order to assess the feasibility of our considerations regarding synthetic targets and initial parallel runs, we investigate three concrete selection strategies. 
Among the most straightforward methods is \textbf{Binary Search (BS)}: 
It recursively compares a target value to the middle element of a sorted value list, and continues searching in either its first or second half. 
The method is very efficient, yet also comparably naive. 
Since a good solution needs to be found after a quick search already, and our objective function is costly to evaluate, the generally more sophisticated \textbf{Bayesian Optimization (BO)} qualifies as method as well. 
We use BO with \emph{Matern5/2} as prior function, and \emph{Expected Improvement (EI)} as acquisition function.
Furthermore, we alter observations, i.e. determined runtimes for investigated CPU limitations, such that they are normalized and turned negative in case of runtime target violations.
This way, we enable BO to better understand pre-defined constraints.
Lastly, we employ a \textbf{Nested Modeling Strategy (NMS)} where our proposed runtime model is directly used for -- given a (synthetic) target runtime -- predicting the next CPU limitation to investigate. 
In the \emph{NMS}, learned model weights are reused for a warm-start of the model training in the next iteration. 
This is possible due to how the individual functions are assembled.

\paragraph{Configuration}
We instantiate our three models, i.e. \emph{BS}, \emph{BO}, and \emph{NMS}, and investigate different settings of sample sizes, synthetic targets, and initial parallel runs. 
Specifically, we use runtimes corresponding to small CPU limitations as synthetic targets.
We determine such CPU limitations as a fraction of $l_{\max}$ using a percentage value $p \in \{0.025, 0.05, \ldots, 0.125, 0.15\}$.
In practice, we make sure to exclude the smallest CPU limitation $0.1$ in order to prevent a prolonging of the overall profiling.
Similarly, we investigate the advantages and disadvantages of $n \in \{2, 3, 4\}$ initial parallel profiling runs, and choose the CPU limitations accordingly.
The complete procedure is outlined in~\autoref{alg:initial_parallel} and applied for all sample size configurations.

\begin{algorithm}
\caption{Initial CPU limitations to profile in parallel}\label{alg:initial_parallel}
\begin{algorithmic}
\Require $p, n, l_{\min}, l_{\max}$
\Ensure $\text{sum}(R_{\text{initial}}) \le l_{\max} \land \vert  R_{\text{initial}} \vert = n$
\State $l_p \gets \max(0.2, l_{\max} \cdot p)$ \Comment{limit of synthetic target}
\State $l_m \gets (l_{\min} + l_{\max}) / 2$ \Comment{middle value}
\State $l_q \gets (l_p + l_{\max}) / 4$ \Comment{approx. quarter value}
\If{$n = 2$}
    \State $R_{\text{initial}} \gets \{l_p, l_{\max} - l_p\}$
\ElsIf{$n = 3 \land l_{\max} > 1$}
    \State $R_{\text{initial}} \gets \{l_p, l_m, l_{\max} - l_m - l_p\}$
\ElsIf{$n = 3 \land l_{\max} \le 1$} \Comment{comfort small CPUs}
    \State $R_{\text{initial}} \gets \{l_p, l_q, l_{\max} / 2\}$
\ElsIf{$n = 4$}
    \State $l_{qm} \gets (l_p + l_q) / 2$ \Comment{compute even smaller value}
    \State $R_{\text{initial}} \gets \{l_p, l_q, l_{qm}, l_{\max} - l_{qm} - l_q - l_p\}$
\EndIf
\State return $R_{\text{initial}}$
\end{algorithmic}
\end{algorithm}

\paragraph{Evaluation Metrics} 
We call into play multiple metrics in order to assess the performance of the various methods to compare. Our primary metric is the Symmetric Mean Absolute Percentage Error (SMAPE), which eases evaluating data sources with different value scales~\cite{hyndman2006another,shcherbakov2013survey,kreinovich2014estimate}. 
The version of SMAPE we utilize provides a result between 0 and 1, which can be interpreted as a percentage value from 0\% to 100\%:
\begin{equation}
    SMAPE = \frac{\sum_{i=1}^n |\hat{Y}_i - Y_i|}{\sum_{i=1}^n (Y_i + \hat{Y}_i)},
\end{equation}
Here,  $Y_i$ denotes the true value and $\hat{Y}_i$ the predicted value.
This is making the assumption that all predicted values are non-negative. 
In practice, we make sure this holds true by altering a model prediction via $\hat{Y}_i=\max(\hat{Y}_i, \epsilon)$, with $\epsilon$ being a sufficiently small non-negative value.

Besides the SMAPE, we furthermore record and evaluate the various methods in terms of
\begin{enumerate}
    \item the number of profiled CPU limitations until a good enough function approximation is possible
    \item the time required for fitting the objective function appropriately, and thus also the associated profiling overhead
\end{enumerate}
With these performance indicators, we compare the methods and carve out their individual characteristics.

\subsection{Results}
We present our findings using the following categories: 

\begin{figure*}[ht]
    \centering
    \includegraphics{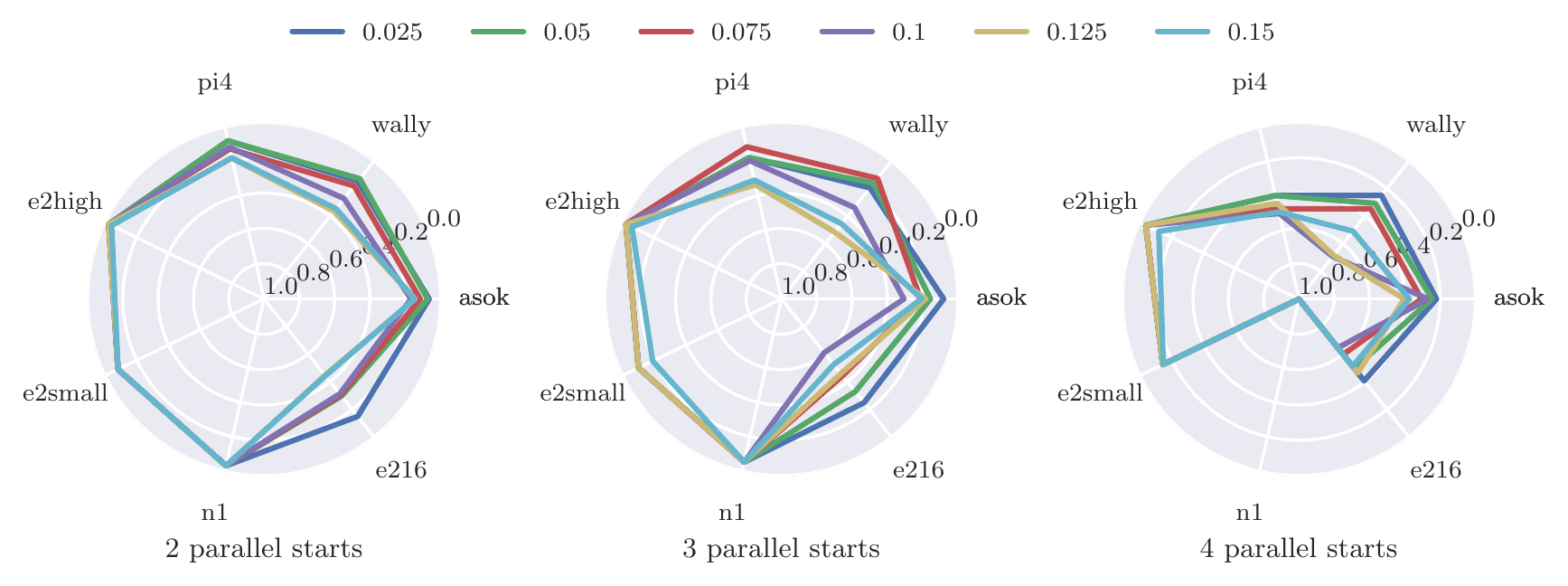}
    \caption{Smallest achievable SMAPE values for different synthetic targets and initial parallel profiling runs.}
    \label{fig:smape_spyder}
\end{figure*}

\subsubsection{Synthetic targets}
In order to evaluate the impact of different configurations of initial parallel profiling runs and synthetic targets, we conducted an experiment in which we tested our profiling approach on datasets generated from all nodes in \autoref{tab:devices}. For this scenario, we are using 10000 profiling samples as it better approximates the presumably exponential behaviour.
\autoref{fig:smape_spyder} depicts the average minimum SMAPE for each algorithm and selection strategy on every node with the given synthetic target configuration.

As expected, a lower synthetic target performs best for machines with a higher amount of CPU cores: This is due to the fact that the exponential increase in execution times takes place at lower CPU limitations and is more or less stable in the higher regions. 
Therefore, to capture this deviation with the selection strategies, the synthetic target limitation should be chosen to be as low as possible.
For instance, the runtime behaviour of the \emph{e216} node with 16 CPU cores can be best fitted with a synthetic target at 2.5\% of all available CPU cores, resulting in the runtime corresponding to $0.4$ CPU millicores. For the \emph{wally} and \emph{asok} nodes with 8 CPU cores, the best results are reached with synthetic targets between 2.5\% and 7.5\%, depending on the number of initial parallel profiling steps.
Furthermore, different synthetic targets thus also result in greater deviations in accuracy for nodes with more CPU cores, which is especially observable at the \emph{e216} node.

As outlined in \autoref{fig:smape_spyder}, on the \emph{e2high}, \emph{e2small} and \emph{n1} nodes with only two and respectively one CPU cores, all synthetic targets perform nearly the same for two parallel initial profiling steps. 
Following~\autoref{alg:initial_parallel}, a limitation of 0.2 CPU millicores was used for each synthetic target between 2.5\% and 10\%. 
The target configurations 12.5\% and 15\% resulted in a limitation of 0.3 CPU for two available cores, indicated by slightly different minimum SMAPE values for the \emph{e2high} and \emph{e2small} nodes.

In addition, the experiment also underlines the motivation to execute the profiling runs directly on the respective devices: Although \emph{e2small} and \emph{e2high} have an identical amount of CPU cores, the same synthetic targets lead to different minimum SMAPE values since the \emph{e2high} virtual machines was equipped with a more powerful CPU. 
Edge and fog computing environments often consists of highly heterogeneous hardware -- i.e. different CPU architectures -- therefore a general profiling only for a specific amount of CPU cores might return vastly different results.

In summary, a configuration of two to three initial parallel profiling runs and a synthetic target between 2.5\% and 7.5\%  on average performed best over all nodes in our experiments. Four initial parallel runs -- although slightly reducing the profiling time -- offer the worst results, especially for low-powered nodes with only one CPU core where four parallel runs are not possible or lightweight single-board devices such as a Raspberry Pi 4.

\subsubsection{Profiling Sample Sizes}
\label{sec:sample_sizes}

\begin{figure}[ht]
    \centering
    \includegraphics{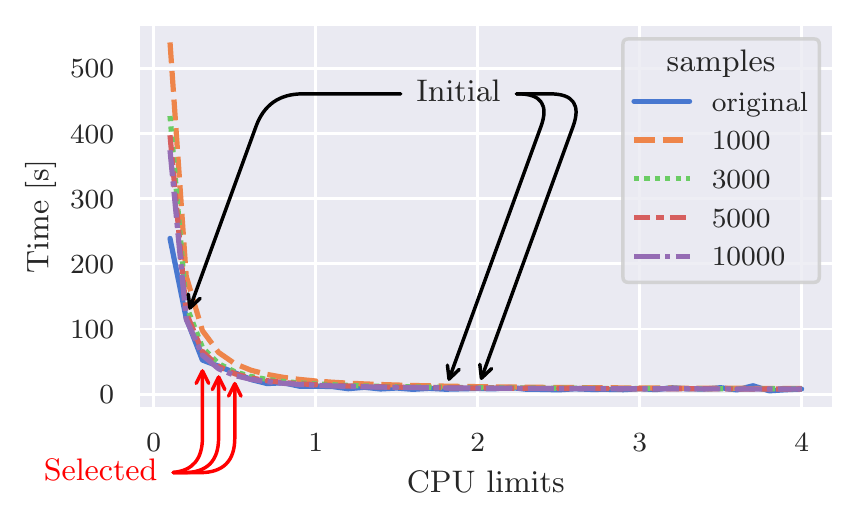}
    \caption{Illustration of NMS choosing profiling points after an initial selection. The number of samples evidently impacts upon the fitting of the curve.}
    \label{fig:increasing_samples_function}
\end{figure}

Whenever a new CPU limitation is selected for profiling a machine learning service, the respective service needs to be provided a suitable amount of test samples in order to approximate its prospective average runtime. 
While more samples generally result in a better estimate, they also impact on the overall time required for profiling. 
It is thus favorable to find a method that works well even with limited sample sizes.
As a consequence, and supplementary to our early stopping approach, we extract the first 1000, 3000, 5000, and 10000 samples of each profiling series respectively, utilize them for computing average runtimes, and eventually use these estimates for evaluating our methods.

\autoref{fig:increasing_samples_function} exemplary depicts the result of the \emph{NMS} selection strategy after six profiled CPU limitations with different amounts of samples for the \emph{Arima} algorithm on the \emph{pi4} node. 
In addition, we configured the profiling with three initial parallel runs and set the synthetic target to 5\% of the available CPU millicores, resulting in the indicated initial profiling points on the plot. 
The selected next profiling points are marked in red and located close to the chosen synthetic target at a CPU limitation of 0.2.

\begin{figure*}[ht]
    \centering
    \includegraphics{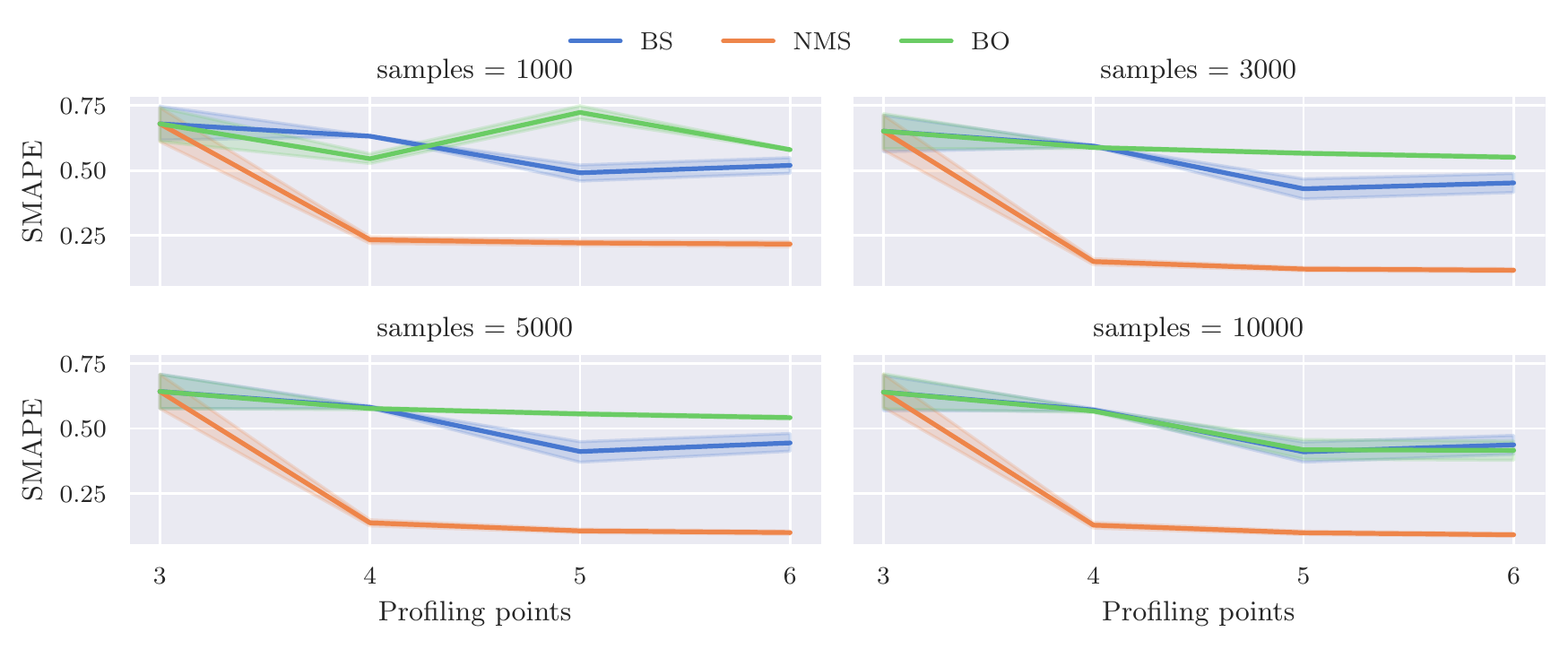}
    \caption{SMAPE across all algorithms on \emph{pi4}, for all sample-size scenarios and an exemplary configuration (3 initial parallel runs, synthetic target = 5\%).}
    \label{fig:smape_pi4_3i_6p}
\end{figure*}

It can be observed that with growing sample sizes, the average runtime per sample can be better approximated, which naturally also impacts upon the fitting capabilities of the employed profiling point selection methods. 
This is also illustrated in~\autoref{fig:smape_pi4_3i_6p} which represents the SMAPE after consecutive profiling steps, for all selection strategies and tested algorithms on the \emph{pi4} node, with three initial parallel profiling steps and the synthetic target at 0.2 CPU millicores (5\%). The Figure also indicates the 95\% confidence interval of all monitored SMAPE values.
As can be seen, the results are relatively stable across the tested algorithms.

\subsubsection{Selection Strategies and Profiling steps}
\autoref{fig:smape_pi4_3i_6p} also depicts the accuracy of different profiling point selection strategies after each profiling step on the aforementioned node. 
All selection strategies start with the same SMAPE, since the first three profiling points are identical. Nevertheless, the accuracy differentiates with consecutive steps based on what points are selected by the respective strategy. 
As can be seen, overall the \emph{NMS} strategy performs best on the \emph{pi4} node, for each configuration of profiling samples, whereas the
\emph{BS} and \emph{BO} strategies result in very similar prediction errors.

The \emph{NMS} selection strategy offers the best results, because it reuses the previously fitted parameters from preceding runtime models: As described in \autoref{sec:approach}, we iteratively replace the runtime function based on the number of available data points, and reuse the previous parameters as initial guesses. 

Finally, it is also observable that all selection strategies already start to converge one to two steps after the initial three parallel profiling runs. 
Consequently, the profiling could be stopped after the fifth step, enabling a faster and more efficient model generation while still offering a nearly optimal result.

\subsubsection{Profiling Time and Accuracy}
\begin{figure*}[ht]
    \centering
    \includegraphics{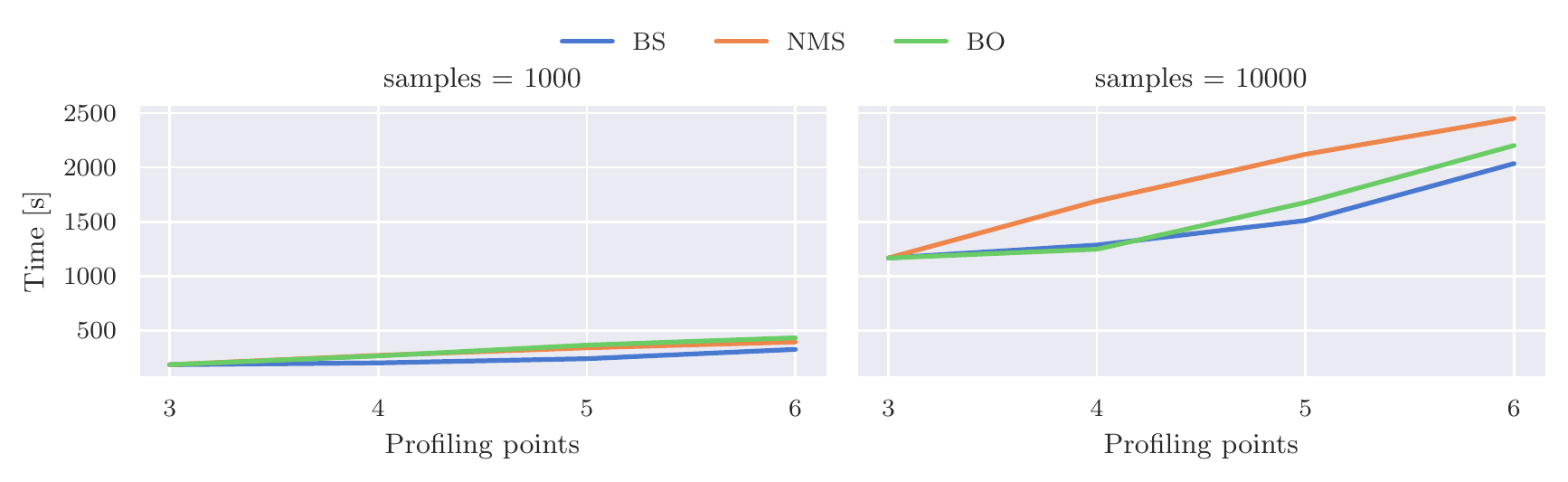}
    \caption{Profiling time required for \emph{Arima} on \emph{pi4}, for two sample-size scenarios and an exemplary configuration (3 initial parallel runs, synthetic target = 5\%).}
    \label{fig:evaluation_strategy_times_trend}
\end{figure*}

Similar to the previous experiment, we also compared the execution time of the implemented selection strategies during succeeding profiling steps. 
As before, we exemplary show in~\autoref{fig:evaluation_strategy_times_trend} the results for the \emph{pi4} node and a configuration with three initial parallel profiling steps as well as a synthetic target of 5\%. 
All algorithms show a very similar trend, thus here we describe the results for the \emph{Arima} algorithm in order to give concrete examples.

It can be observed that naturally, the execution times are higher for larger sample sizes, since the individual profiling steps require more time. 
Yet, the various methods show a similar trend for increasing profiling steps: Depending on which CPU limitations were chosen by the selection strategies, the execution time increases more or less linearly. 
With succeeding profiling steps, the \emph{NMS} strategy takes slightly more time than \emph{BS} and \emph{BO} which remain comparably fast.
This is due to the fact, that while the \emph{NMS} method already chooses a profiling point expected to be closer to the target, the other strategies approach the synthetic target either starting from higher CPU limitations (\emph{BS}) or initially lack a strong prior belief (\emph{BO}). 

Although it takes longer to execute a profiling run for a smaller CPU limitation
(i.e. 268s for four profiling steps in case of \emph{NMS} and respectively 199s for \emph{BS} and 263s for \emph{BO} when using 1000 samples), the overall runtime behaviour can already be fitted significantly better with \emph{NMS}. 
As also indicated by the trend in \autoref{fig:smape_pi4_3i_6p}, for 1000 samples and the \emph{NMS} strategy, the SMAPE is already decreased to only 0.29 after the fourth profiling step, 
whereas \emph{BS} and \emph{BO} still result in 0.62 and 0.38.

Compared to four steps, when using the \emph{NMS} method, the maximum amount of six profiling steps for \emph{Arima} on \emph{pi4} increases the execution time by around 45\% from 268 to 392 seconds (1000 samples) or respectively from 1690 to 2451 seconds (10000 samples), while only optimizing the SMAPE from 0.29 to 0.27 (1000 samples) or 0.18 to 0.11 (10000 samples). In case of \emph{BS} and \emph{BO}, the SMAPE does not change significantly for higher amounts of profiling samples, while at the same time the execution time is increased by between 58\% and 76\%.

Consequently, the results show that the execution time can be significantly decreased by stopping the profiling after only four to five steps, since the SMAPE is only slightly improved afterwards. In regards to the sample size configuration, for all selection strategies the SMAPE is decreased by up to 0.15 when using 10000 samples instead of 1000, however the profiling takes about five times longer.

Therefore, in case the profiling should be conducted as fast as possible, it makes sense to use less samples and profiling steps since the accuracy is only slightly improved during longer profiling.

Finally, to further increase the confidence in the values used for the fitting while still decreasing the overall profiling time, the early stopping approach described in \autoref{sec:earlystopping} can be used: For the aforementioned scenario, a 95\% confidence level and early stopping criteria of 10\%, the overall profiling time amounts to 1135 seconds and decreases the SMAPE to 0.13 for six total profiling steps.
Consequently, the early stopping method decreases the profiling time by around 50\% while still achieving a similar accuracy to 10000 samples.

\subsubsection{General accuracy of selection strategies}
As the final evaluation, we conducted an experiment in which we evaluated the average SMAPE for profiling steps in the range between four and eight across all nodes in~\autoref{tab:devices} and all tested algorithms. 
In  this scenario, we started with three initial parallel profiling runs and configured the method to use 10000 profiling samples. In order to further strengthen our evaluation, we also provided a \emph{Random} selection strategy which randomly chooses profiling points after the initial parallel ones. The experiment was repeated 50 times.

\begin{figure}
    \centering
    \includegraphics{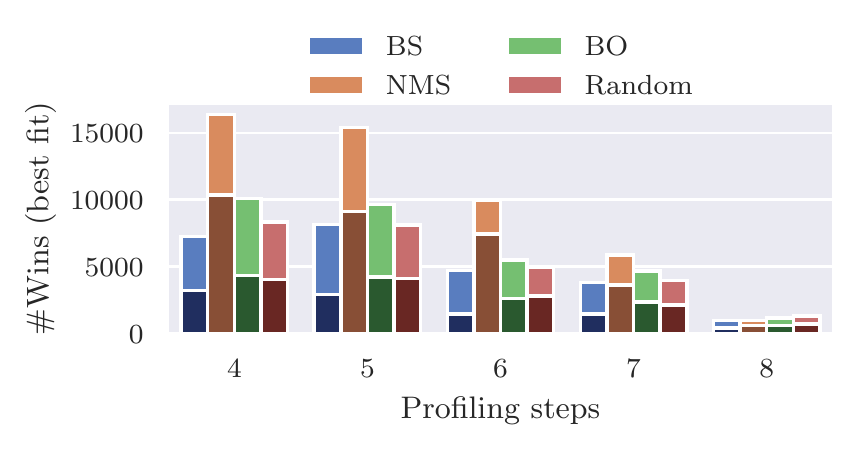}
    \caption{Number of wins for each strategy and number of profiling steps. A win is indicated by being (among) the producer(s) of smallest prediction errors. The dark (0\%) and light bars (10\%) depict different tolerance policies.}
    \label{fig:smape_selection_strategies_winner}
\end{figure}

\autoref{fig:smape_selection_strategies_winner} shows the distribution of best-performing selection strategies with the given amount of profiling points for all evaluation nodes and algorithms. Morover, we also count (near-) winning strategies in a 10\% range and depict them using the light colored bars.

As indicated by the results, the NMS approach is able to outperform the other selection methods over all nodes, especially for smaller amounts of profiling steps. In some cases, \emph{Random} is able to offer a similar performance to \emph{BO} and \emph{BS}, since it uses the same three initial points and with fewer available CPU cores the likelihood of choosing suitable points increases. At the same time,
less profiling steps are needed for fewer CPU cores, resulting in decreasing winning strategies for more profiling steps.

\section{Related Work}
\label{sec:related_work}

In the area of resource management, runtime behavior modeling of tasks is often applied to scale-out jobs in order to comply with given timing requirements. 
For instance, Bell~\cite{thamsenSelectingResourcesDistributed2016} leverages historical data about previous dataflow job executions to model the scale-out behaviour and subsequently choose suitable resource configurations with regards to runtime targets. 
In following works the method was extended to support dynamic scaling of dataflow workloads~\cite{scheinertEnelContextAwareDynamic2021} and similar to our approach a binary search method is applied when not enough historical training data is available~\cite{thamsenEllisDynamicallyScaling2017a}. However, in contrast to our approach the aforementioned proposals rely on user specified runtime targets and target distributed dataflow systems in cloud environments, where jobs consists of several tasks, often organized as directed acyclic graphs.
We aim to create global runtime models directly on heterogeneous devices, without specified targets.

Lightweight virtualization technologies enable a more fine-grained approach to resource management and are especially suitable for edge computing environments~\cite{al-rakhamiLightweightCostEffective2020a}.
Container orchestration stacks allow for vertical and horizontal scaling of the application containers and several approaches extend them in order to enable an adaptive autoscaling.
Libra~\cite{ballaAdaptiveScalingKubernetes2020} combines vertical and horizontal pod scaling by monitoring the CPU utilization: First, a sufficient CPU limitation for the current workload is found during vertical up-scaling before more replicas are spawned to distribute the load.
Similarly, the ElasticDocker method~\cite{al-dhuraibiAutonomicVerticalElasticity2017} also vertically scales application containers during increasing workload and in case the current node does not offer enough resources, live migrates the container to another node.
Yet, many works consider autoscaling from the perspective of current resource utilization.
In our approach we aim to scale on -- and therefore dynamically adapt to -- changing frequencies in sensor data streams to enable a just-in-time computation of incoming data samples and subsequently improve the environmental and self-awareness of devices in the Cognitive Cloud Continuum \cite{ferrerCognitiveComputeContinuum2021a}.

Although there are approaches which take stream arrival rates into account \cite{dematteisElasticScalingDistributed2017, hoseinyfarahabadyQFlinkQoSAwareController2020},
they often apply model predictive control principles to predict upcoming load intensities and scale the parallelism of applied streaming operators. The prediction models are either trained on historical data from completed containers \cite{buchacaProactiveContainerAutoscaling2020} or by feedback loop based system which rely on re-calibration processes in case of performance degradation \cite{liuStepwiseAutoProfilingMethod2018}. In contrast, we focus on a single and efficient profiling process which enables the adaptive adjustment of resources based on the initially created model.

\section{Conclusion}
\label{sec:conclusion}
This paper presented our approach to efficient runtime profiling for black-box machine learning services, which is particularly important in edge computing environments, where constrained resources and changing workloads are typically limiting factors.
We propose the utilization of synthetic runtime targets during profiling for building general runtime models.
In order to obtain appropriate runtime models with few samples only, we motivate a nested modeling strategy, which effectively leads to a fastly converging profiling phase. 
Furthermore, we initially conduct multiple profiling runs in parallel in order to make the most out of available resources. 

We conducted an exhaustive evaluation of our approach, in the process implementing and comparing multiple concrete selection strategies and carving out their individual advantages and disadvantages.
As shown by our results, the proposed nested modeling strategy exhibits a clearly superior performance in terms of prediction accuracy, while requiring only negligible more time than comparative methods. 
The advantage of our method is especially significant in cases of limited data availability or time-constrained profiling. 

In the future, we plan to integrate our approach directly into lightweight container orchestration platforms such as KubeEdge. 
Moreover, we want to investigate how our findings with regards to profiling can be transferred to related domains other than stream-based sensor analytics, for example distributed dataflows or scientific workflows.

\bibliographystyle{IEEEtran}
\bibliography{references}

\end{document}